\documentclass[showpacs,superscriptaddress,floatfix,twocolumn,amsmath,amssymb,prb]{revtex4-1}
\usepackage{dcolumn}
\usepackage{yfonts}
\usepackage{bm}
\usepackage{graphicx}
\usepackage{color}
\usepackage[normalem]{ulem}

\newcommand{\Ket}[1]{\vert #1 \rangle}
\newcommand{\Bra}[1]{\langle #1 \vert}
\newcommand{\MatEl}[3]{\langle #1 \vert #2 \vert #3 \rangle}
\newcommand{\n}{\mbox{\boldmath $\nabla$}}

\newcommand{\nbar}{\bar n}

\def\hb0{h_{\rm b}^{(0)}}
\def\p12{p_{12}({\bf q},t)}

\begin{document}

\title{Vibration multistability and quantum switching for dispersive coupling}

\author{ Z. Maizelis}
\affiliation{A. Ya. Usikov Institute for Radiophysics and Electronics,
National Academy of Sciences of Ukraine, 61085 Kharkov, Ukraine}
\author{M. Rudner}
\affiliation{Niels Bohr International Academy and the Center for Quantum Devices, Niels Bohr Institute, University of Copenhagen, 2100 Copenhagen, Denmark}
\author{M. I. Dykman}
\affiliation{Department of Physics and Astronomy, Michigan State University, East Lansing, MI 48824, USA}

\date{\today}

\begin{abstract}

We investigate a resonantly modulated harmonic mode, dispersively coupled to a nonequilibrium few-level quantum system. We focus on the regime where the relaxation rate of the system greatly exceeds that of the mode, and develop a quantum adiabatic approach for analyzing the dynamics.  Semiclassically, the dispersive coupling leads to a mutual tuning of the mode and system into and out of resonance with their modulating fields, leading to multistability. In the important case where the system has two energy levels and is excited near resonance, the compound system can have up to three metastable states. Nonadiabatic quantum fluctuations associated with spontaneous transitions in the few-level system lead to switching between the metastable states. We provide parameter estimates for currently available systems. 
\end{abstract}

  \maketitle

\section{Introduction}
\label{sec:Introduction}

Dispersive coupling of a quantum system to a mechanical or electromagnetic cavity mode has been attracting much attention recently. The coupling provides a means for quantum nondemolition measurement of the occupation number of the mode or of the populations in the energy levels of the system \cite{Schuster2005,Haroche2006,Schuster2007,Thompson2008,Vijay2009,LaHaye2009,Clerk2010,Petersson2012}. The underlying read-out mechanism is the shift of the mode frequency or the system transition frequency, which depends on the state populations of the system or the mode, respectively. In the dispersive regime, a measurement erases information about the quantum phase, but does not cause transitions between energy levels. However, such transitions can happen due to coupling to a thermal reservoir, and also if the mode and/or the system are modulated by external fields. It is well understood that, through dispersive coupling, thermal interstate transitions cause decoherence \cite{Schuster2005,Eichenfield2009,Weiss2013}. Much less is known about the effects of periodic modulation and the interplay of the modulation and dephasing due to the coupling to a thermal reservoir. 

In this paper we address these problems. We consider a mode $\mathcal{ M}$ (a harmonic oscillator) coupled to a dynamical system $\mathcal{S}$. The mode and the system are also coupled to separate thermal reservoirs and can be modulated by periodic fields. The couplings and the modulation are assumed weak in the sense that the coupling energy is small compared to the interlevel energy spacing. In other words, the widths of the energy levels and the Rabi energies are small compared to the level spacing. The modulation is assumed to be nearly resonant and will be described in the rotating wave approximation (RWA). 

In distinction from the celebrated Jaynes-Cummings model \cite{Mandel1995,Walls2008,Carmichael2008}, here the level spacings of the mode and the system are significantly different. For a dispersive ${\mathcal M} -{\mathcal S}$ coupling the major effect is not energy exchange, but rather it is the change of the level spacing depending on the state population, which occurs already in the first order in the coupling constant. Semiclassically, this situation can give rise to multistability in the response to a modulating field as follows.

\begin{figure}[ht]
\includegraphics[width=7.0truecm]{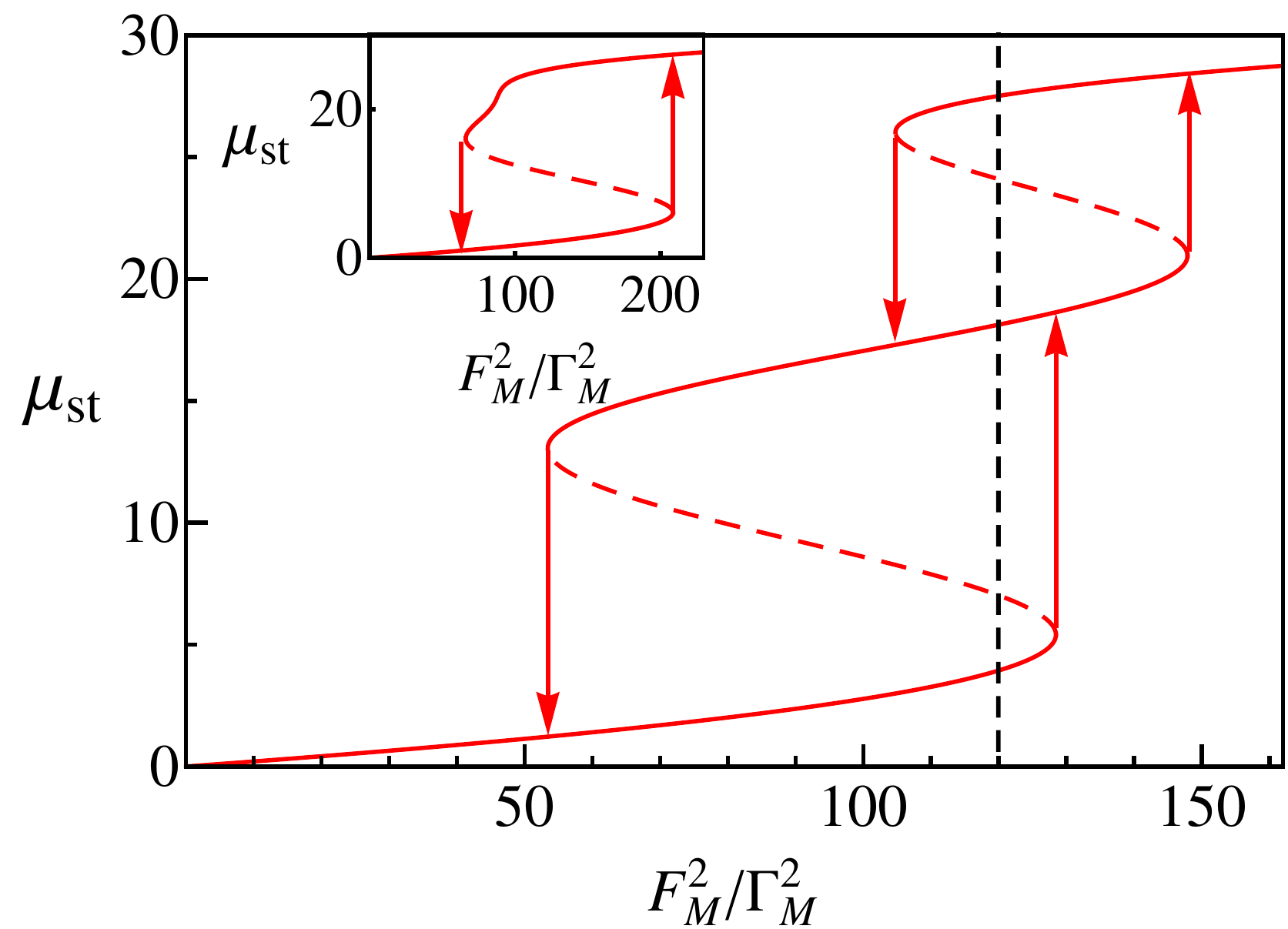}
\caption{Tristability of a modulated 
mode dispersively coupled to a two-level system. The ordinate gives the stationary occupation number $\mu_{\rm st}$ of the mode (the squared mode amplitude is $ 2\hbar \mu_{\rm st}/m_{_M}\omega_{_M}$, where $m_{_M}$ and $\omega_{_M}$ are the mode mass and frequency, respectively). The abscissa shows the reduced squared modulation amplitude $F_{_M}^2$. Both the mode and the system are resonantly modulated. In the rotating wave approximation, the model is described by Eqs.~(\ref{eq:H_RWA1}), (\ref{eq:Liouville_osc}) and (\ref{eq:Bloch}). The ratio of the relaxation rate of the two-level system $\Gamma_{_S}\equiv (2\tau_{_S})^{-1}$ to the relaxation rate of the mode $\Gamma_{_M}\equiv \tau_{_M}^{-1}$ is 30; the reduced amplitude of the field modulating the two-level system is $F_{_S}/\Gamma_{_S}=24$. The reduced detunings of the modulating fields from the transitions frequencies of the mode and the system are, respectively, $(\omega_{F_M}-\omega_{_M})/\Gamma_{_M}=1$ and $(\omega_{F_S}-\omega_{_S})/\Gamma_{_S}= -10$. The reduced strength   of the dispersive coupling is $V/\Gamma_{_M}=15$. The inset refers to $(\omega_{F_M}-\omega_{_M})/\Gamma_{_M}=0.3$, in which case the system does not show tristability. The unstable states are shown by dashed lines. The black vertical dashed line shows the modulation field used in Fig.~\protect\ref{fig:graphic_solution} below.}
\label{fig:hysteresis}
\end{figure}

For given modulating field parameters, the combined system may self-consistently support either large amplitude forced vibrations of mode $\mathcal{M}$, with the effective mode frequency tuned into good resonance with the driving field via the dispersive coupling, or small amplitude vibrations with an effective mode frequency far from resonance with the driving field. In each case, the vibration amplitude of mode $\mathcal{M}$ sets the transition frequencies of the system $\mathcal{S}$. If system $\mathcal{S}$ is modulated itself, this determines its quasi-steady-state level occupations. Through the dispersive coupling, these level occupations tune the oscillator frequency into or out of resonance with the driving field, leading to the self-consistent mean-field multi-stability, see Fig.~\ref{fig:hysteresis}.

The mean-field theory describes the semiclassical multistability of the system, but does not account for the role of {\it fluctuations}. Classical and quantum fluctuations unavoidably come along with relaxation as a consequence of coupling to a bath. 
In multistable systems, fluctuations cause interstate switching, {\it even at zero temperature}. 

Interestingly, where the dispersive coupling is weak, there is obviously no multistability; however, where it is strong there is also no multistability, because the switching rate becomes comparable to the relaxation rate, and then the very notion of multistability becomes meaningless. In what follows we find the appropriate range of the coupling strength. We provide a general formulation of a mean-field theory and a theory of the switching rates in the important case where the typical relaxation time $\tau_{_S}$ of the system $\mathcal{S}$ is much smaller than the relaxation time $\tau_{_M}$ of the mode. 

\subsection{The rotating wave approximation}

Formally, the dispersive coupling Hamiltonian $H_i(\hat{M},\hat{S})$ is a function of a mode operator $\hat{M}$ and a system operator $\hat{S}$, which commute with the isolated mode and system Hamiltonians, respectively, in the absence of modulation. 
For example, $\hat{M}$ can be the occupation number of the mode, $a^\dagger a$, where $a$ and $a^\dagger$ are the lowering and raising operators, while, if the dynamical system is a spin in a static magnetic field $B_z$, $\hat {S}$ can be the spin operator $s_z$. This form of coupling assures that $H_i$ is independent of time in the interaction representation.

For illustration, we consider a mode coupled to a two level system (TLS), each modulated by its own nearly-resonant field (with $\hbar = 1$):
\begin{eqnarray}
\label{eq:osc+spin}
H_{_S}
&=& \omega_{_S} s_z - s_xF_{_S}\cos\omega_{F_S}t, \nonumber\\
H_{_M} 
&=& \omega_{_M}a^\dagger a  - (a+a^\dagger)F_{_M}\cos\omega_{F_M}t.
\end{eqnarray}
Here $s_{x,z} = \sigma_{x,z}/2$, where $\sigma_{x,z}$ are Pauli operators which act on the TLS. For nearly-resonant modulations, the detunings  $\delta\omega_{_M}=\omega_{F_M}-\omega_{_M}$ and $\delta\omega_{_S}=\omega_{F_S}-\omega_{_S}$  of the modulation frequencies from the transition frequencies $\omega_{_M}$ and $\omega_{_S}$ are small compared to the transition frequencies themselves, and to their difference:  $|\delta\omega_{_M}|, |\delta\omega_{_S}|\ll \omega_{_M}, \omega_{_S}, |\omega_{_M}-\omega_{_S}|$. The condition on $|\omega_{_M}-\omega_{_S}|$ in particular justifies the approximation where only dispersive coupling is taken into consideration.

The simplest form of the dispersive coupling of a mode and a TLS is $H_i= Va^\dagger a s_z$, which we now consider. We switch to the interaction representation using the unitary transformation $U(t)=\exp(-i\omega_{F_M}a^\dagger a t - i\omega_{F_S}s_z t)$. Disregarding the fast-oscillating (counter-rotating) terms proportional to the modulation amplitudes $F_{_S},F_{_M}$, in the spirit of the RWA, we write the transformed Hamiltonian ${\tilde H} = \left[U^\dagger(t)\left({H}_{_M} + {H}_{_S} + {H}_i\right)U(t)-iU^\dagger (t) {\dot U}(t)\right]_{RWA}$ as 
\begin{equation}
\label{eq:Hamiltonian_rotating_frame}
\tilde{H} = \tilde{H}_{_M} + \tilde{H}_{_S} + \tilde{H}_i,
\end{equation}
with
\begin{eqnarray}
\label{eq:H_RWA1}
{\tilde H}_{_M} &=& -\delta\omega_{_M} a^\dagger a -\frac{1}{2}F_{_M}(a+a^\dagger),\\
\nonumber \qquad {\tilde H}_{_S} &=& -\delta\omega_{_S} s_z - \frac{1}{4}F_{_S}(s_+ + s_-),\\
\nonumber \tilde H_i &=& Va^\dagger a s_z.
\end{eqnarray}
Model (\ref{eq:H_RWA1}) describes, in particular, the dispersive coupling of a cavity mode to a two-level atom in cavity QED or to an effectively two-level Josephson junction in circuit QED, which has been studied in many experiments, see e.g.~Refs.~\onlinecite{Wallraff2004,Schuster2007,Bertet2012} and references therein. More generally, $H_i$ may take on a more complicated form. In particular, the coupling does not have to be linear in $a^\dagger a$. Similarly, when system $\mathcal{S}$ has more than two levels, the coupling Hamiltonian may involve more complicated combinations of system operators as well.
We will generally characterize the energy of dispersive coupling by a parameter $V$, even where the coupling has a form different from $\tilde H_i$ in Eq.~(\ref{eq:H_RWA1}); we assume $|V|\ll \omega_{_S},\omega_{_M},|\omega_{_S}-\omega_{_M}|$.

\subsection{Master equation}
\label{subsec:intro_master_equation}

In order to describe the dynamics in the presence of dissipation, we consider the density matrix $\rho$ of the coupled mode and system. Assuming Markovian dynamics in slow time, i.e.~on times long compared to $\omega_{_M}^{-1},\omega_{_S}^{-1}, |\omega_{_M}-\omega_{_S}|^{-1}$, we can write the equation of motion for $\rho$ in the interaction representation in the form:
\begin{equation}
\label{eq:rho_{_S}chematic}
\dot \rho = \hat L \rho  \equiv \hat L_{_M}\rho + \hat L_{_S}\rho + i [\rho,  \tilde H_i].
\end{equation}
Here, $\hat L_{_M}$ and $\hat L_{_S}$ are Liouville operators, or superoperators, cf.~Ref.~\onlinecite{Zwanzig2001}; they describe, respectively, the dynamics of the mode and the system coupled to their thermal reservoirs but isolated from each other.

Below we will calculate the density matrix in the basis where operators $\hat {M}$ and $\hat {S}$ are diagonal.  
Importantly, $\rho$ must remain Hermitian through its evolution via Eq.~(\ref{eq:rho_{_S}chematic}).  
As a consequence, for any operator of the mode and the system $\hat O_{MS}$,
\begin{equation}
\label{eq:hermitivity}
 (\hat L \hat O_{MS})^\dagger = \hat L \hat O_{MS}^\dagger.
\end{equation}
This condition applies also to $\hat L_{_M}$ and $\hat L_{_S}$ taken separately.

In the frequently used model of dissipation where coupling of the mode to a thermal reservoir is taken to be linear in the operators $a,a^\dagger$, to the leading order in this coupling we have\cite{Mandel1995}:
\begin{eqnarray}
\label{eq:Liouville_osc}
\hat L_{_M}\rho &=& - \Gamma_{_M}\left [(\nbar + 1)(a^\dagger a \rho - 2 a\rho a^\dagger + \rho a^\dagger a)\right. \nonumber\\
&&\left.+ \nbar  (aa^\dagger \rho - 2 a^\dagger\rho a + \rho aa^\dagger)\right] + i[\rho,{\tilde H}_{_M}],
\end{eqnarray}
where $\nbar \equiv\nbar(\omega_{_M}), \nbar(\omega)=[\exp(\omega/k_BT)-1]^{-1}$ is the mode Planck number and $\Gamma_{_M}$ is the decay rate. 
We note that Eq.~(\ref{eq:Liouville_osc}) is not limited to describing Ohmic dissipation; in the microscopic derivation it is assumed that $\Gamma_{_M}\ll \omega_{_M}$ and $|d\Gamma_{_M}/d\omega_{_M}|\ll 1$, and that the time is slow, cf.~Ref.~\onlinecite{DK_review84}. We assume that the renormalization of the parameters of the mode due to the coupling to the thermal reservoir has been incorporated into the parameter values.

A simple form of relaxation for the two-level system is described 
via Bloch equations. 
In this case operator $\hat L_{_S}\rho$ in Eq.~(\ref{eq:rho_{_S}chematic}) has the same form as $\hat L_{_M}\rho$, except that (i) the friction coefficient $\Gamma_{_M}$ should be replaced by the parameter $\Gamma_{_S}$ that gives the reciprocal lifetime of the two-level system, $\tau_{_S}^{-1}=2\Gamma_{_S}[2\bar n(\omega_{_S})+1]$, where $\bar n_{_S} = \bar n(\omega_{_S})$ is the Planck number; (ii)  operators $a$ and $a^\dagger$ in the dissipation term should be replaced by $s_-$ and $s_+$, respectively, with $s_\pm = s_x\pm is_y$, and (iii) Hamiltonian $\tilde H_{_M}$ should be replaced with $\tilde H_{_S}$. 
Further, we incorporate additional trasverse relaxation 
through a term $-\Gamma_{\perp}(\rho - 4 s_z\rho s_z)/2$ in $\hat L_s\rho$.

\subsection{Multistability in a simple model of dispersive coupling}
\label{subsec:intro_multistability}

To build intuition before the more technical discussion, we now provide a heuristic semi-quantitative picture of the adiabatic mean-field multistability for dispersive coupling to a TLS; the justification and the applicability conditions follow from the general analysis in Sec.~\ref{sec:mean_field} below. Suppose that the mode is in a state $\Ket{m}$ with $\MatEl{m}{\tilde{H}_i}{m} = V m s_z$.
For the mode in this state, the detuning of the effective TLS transition frequency from the driving field frequency is given by $\delta\omega_{_S}(m)=\delta\omega_{_S} -Vm$, as seen from Eqs.~(\ref{eq:Hamiltonian_rotating_frame}) and (\ref{eq:H_RWA1}). In the adiabatic approximation we solve for the dynamics of the two-level system assuming that this frequency detuning is independent of time. Using the well-known result for this problem, see e.g.~Ref.~\onlinecite{Karplus1948}, we obtain the mean value of $s_z$ for a given value of $m$:
\begin{eqnarray}
\label{eq:Bloch}
&&\langle s_z\rangle_{_S} = -\Gamma_{_S} \left[2\Gamma_{_S}(2\bar n_{_S} +1)+\frac{1}{4}\frac{\gamma F_{_S}^2}{\gamma^2+\delta\omega_{_S}(m)^2}\right]^{-1}\nonumber\\
&&\gamma = \Gamma_{_S}(2\bar n_{_S}+1) + \Gamma_{\perp}, \quad \delta\omega_{_S}(m)=\delta\omega_{_S} -Vm,
\end{eqnarray}
where $\gamma$ is the decay rate of the spin components $s_\pm$.

Through the interaction term $\tilde{H}_i$ in Eqs.~(\ref{eq:Hamiltonian_rotating_frame}) and (\ref{eq:H_RWA1}), the average TLS population difference $\langle s_z\rangle_{_S}$ acts back on the mode, changing its frequency by $\nu(m)\equiv V\langle s_z\rangle_{_S}$. Importantly, the mode frequency depends on its degree of excitation, $m$. Such dependence is characteristic for nonlinear modes. Here it comes from the resonant pumping of the TLS. In turn, the typical values of $m$ in the stable 
mode state determine the detuning of the TLS from the forcing $F_{_S}\cos\omega_{F_S}t$ that modulates it, $\delta\omega_{_S}(m)\equiv \omega_{F_S}-\omega_{_S}- Vm$, thus determining $\langle s_z\rangle_{_S}$. 

The mutual tuning of the mode and the two-level system to resonance leads to multistability of the compound system. Indeed, the 
stationary state mean occupation number of a resonantly modulated harmonic oscillator is given by the familiar expression $m_{\rm st} = \frac{1}{4}F_{_M}^2/[\Gamma_{_M}^2+ (\omega_{F_M}-\omega_{_M})^2]$. Given the dependence of the mode frequency on its degree of excitation, $m$, one might expect to find a self-consistency relation for the stationary state of the form:
\begin{equation}
\label{eq:naive_m}
m_{\rm st} =  \frac{1}{4}F_{_M}^2/\{\Gamma_{_M}^2+ (\omega_{F_M}-\omega_{_M}-\nu(m_{\rm st})]^2\}.
\end{equation}
The resulting system of nonlinear equations (\ref{eq:Bloch}) and (\ref{eq:naive_m}) can have multiple solutions. An example is the dependence of the squared mode vibration amplitude (equal to $2\hbar m_{\rm st}/\omega_{_M}$, for a unit mode mass)
on the modulation strength, which is shown in Fig.~\ref{fig:hysteresis}. In fact, the quantity plotted is the mean-field value of the ``center of mass'' variable $\mu_{\rm st}$ of the quasi-stationary Wigner distribution over the occupation numbers $m$ of the mode; it is given by Eq.~(\ref{eq:stationary_states}), which for the considered model coincides with Eq.~(\ref{eq:naive_m}) and justifies the above qualitative arguments.

For the chosen parameters the mode can have up to three stable states at a time. In the mean-field picture where quantum and classical fluctuations are neglected (see Sec.~\ref{sec:fluctuations} for the role of fluctuations), this tristability is revealed by
a hysteresis pattern with multiple switching between stable branches with the varying control parameter (here, the driving strength).

\begin{figure}[ht]
\includegraphics[width=7.0truecm]{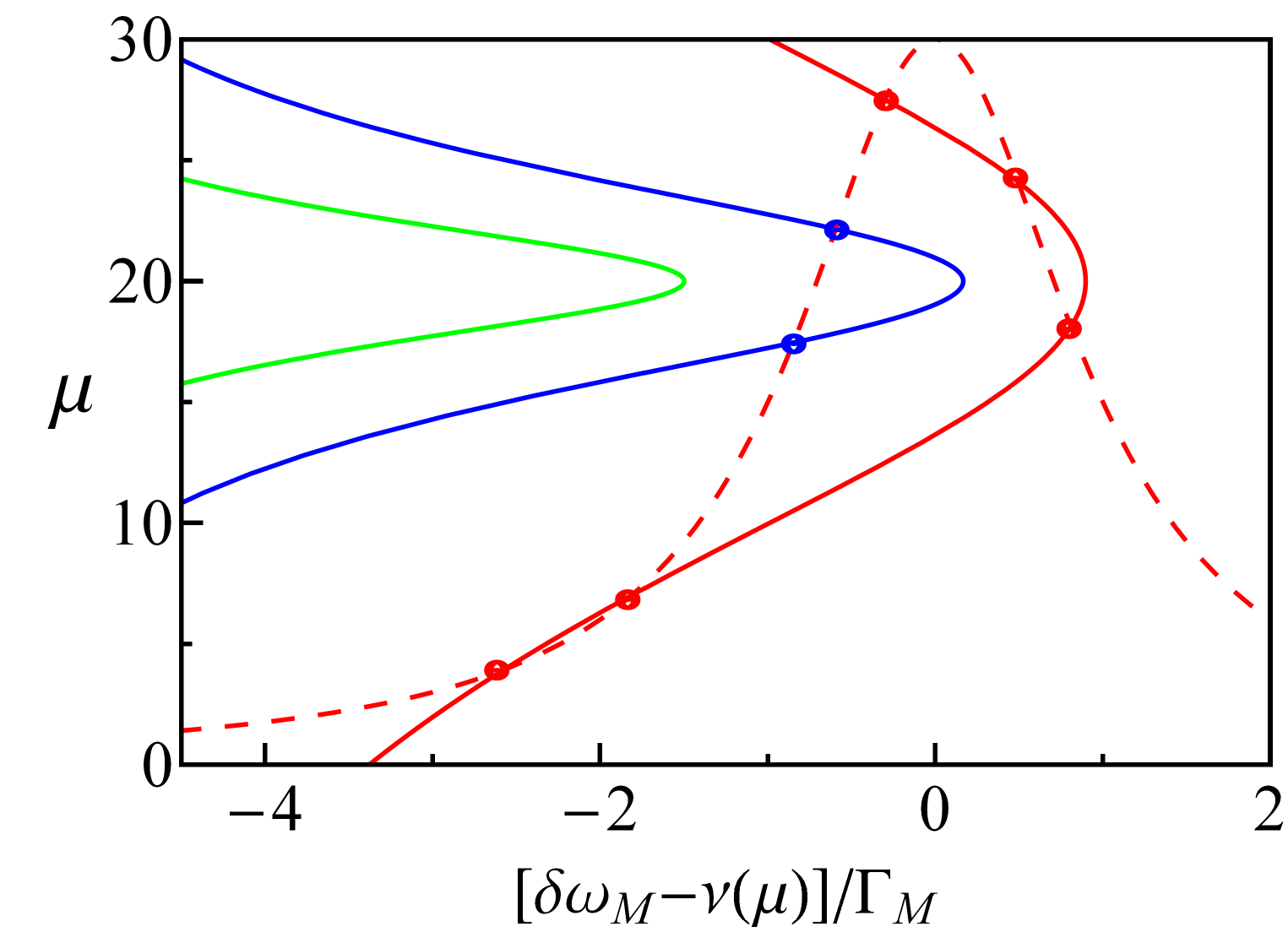}
\caption{Onset of multistability for dispersive coupling. The solid lines show the change of the population difference of the two-level system $\langle s_z\rangle_{_S}$ as a function of the average occupation number of the mode $\mu=\langle m\rangle$; for convenience, instead of $\langle s_z\rangle_{_S}$  we plot the reduced frequency shift $-\nu(\mu)=-V\langle s_z\rangle_{_S}$  counted off from $\delta\omega_{_M}$. The  dependence of $\langle s_z\rangle_{_S}$ on $\mu$ is resonant, which corresponds to the tuning of the two-level system in resonance with the modulating field by varying the mode occupation number. The blue, green, and red solid curves correspond to $F_{_S}/\Gamma_{_S} = 4, 8$ and 24 in Eq.~(\ref{eq:Bloch}). Other parameters are the same as in Fig.~\ref{fig:hysteresis}. The dashed line shows the resonant response $\mu\equiv \langle m\rangle$ of a {\em linear} mode with eigenfrequency $\omega_{_M}+\nu(\mu)$, cf. Eq.~(\ref{eq:naive_m}); here we use $\nu(\mu)$ as a variable and find $\mu$ for a given $\nu(\mu)$ and for $F_{_M}^2/\Gamma_{_M}^2=120$; such $F_{_M}$ corresponds to the long-dash vertical line in Fig.~\ref{fig:hysteresis}. The points show the solutions of Eqs.~(\ref{eq:Bloch}) and (\ref{eq:naive_m}); the small-$\mu$ intersections of the dashed line with the blue and green solid lines occur for smaller $\delta\omega_{_M}-\nu(\mu)$ and are not shown.}
\label{fig:graphic_solution}
\end{figure}

The onset of multistability can be understood from the graphical solution of Eqs.~(\ref{eq:Bloch}) and (\ref{eq:naive_m}), illustrated in Fig.~\ref{fig:graphic_solution}. The solid lines on this figure show the resonant dependence of the reduced population difference of the TLS (see caption) on the ``center of mass'' occupation number of the mode $\mu$. It is given by Eq.~(\ref{eq:Bloch}) with $m$ replaced by $\mu$. The resonance is a consequence of the TLS frequency detuning $\delta\omega_{_S}(\mu)$ being linear in $\mu$. The dashed line shows the resonant dependence of the scaled squared amplitude of the modulated mode $\mu$ on the mode frequency. Note that there is always an odd number of intersections; for the blue and green curves it is equal to 1 and the intersection occurs for small $\mu$ outside the range shown in the figure. The corresponding regime corresponds to the single stable state of the modulated compound system. The case of 3 intersections corresponds to bistability, whereas 5 intersections correspond to tristability. The understanding of this pattern comes from the analysis of the bifurcation curves in Sec.\ref{subsec:multistability}.

The possibility of bistability of the response of a mechanical mode to resonant modulation in the situation where the mode is coupled to another system (a massive classical particle diffusing along the mechanical resonator) was considered earlier \cite{Atalaya2011a}. Such coupling is similar to dispersive coupling, as the diffusion changes the mode frequency and is in turn affected by the vibrations. However, in contrast to Ref.~\onlinecite{Atalaya2011a}, the analysis below is fully quantum, it is general as it is not limited to a specific coupling mechanism, the mean-field predictions are different (for example, tristability), and most importantly, the class of systems to which the results refer is much broader.

The rest of the paper is organized as follows.
In Sec.~\ref{sec:adiabatic} we develop equations of motion for the mode density matrix, and 
introduce the adiabatic approximation which allows a dominant and tractable part of the coupled set of equations to be isolated.
Then in Sec.~\ref{sec:mean_field} we consider the semiclassical limit of large mode vibration amplitude, and derive mean field equations which govern the stationary state vibration amplitudes and phases of the mode. The mean field equations capture the multistability of the system and its critical slowing-down near bifurcation points in parameter space. In Sec.~\ref{sec:fluctuations} we study fluctuations and switching between the mean-field metastable states, induced by random transitions of system $\mathcal{S}$ through the $\mathcal{M}-\mathcal{S}$ coupling. Finally, in Sec.~\ref{sec:conclusions} we summarize our main conclusions and discuss the relevance for various experimental systems of current interest.

\section{Adiabatic approximation}
\label{sec:adiabatic}

The central assumption of our analysis is that the relaxation time $\tau_{_S}$ of system $\mathcal{S}$ is much smaller than the relaxation time $\tau_{_M}$ of the mode [which is given by  $\Gamma_{_M}^{-1}$ for the model (\ref{eq:Liouville_osc})]. 
We exploit this separation of timescales to solve Eq.~(\ref{eq:rho_{_S}chematic}) in an adiabatic approximation: first we solve for the evolution of system $\mathcal{S}$ for a fixed state of the mode, and then examine how the resulting quasi-stationary states of $\mathcal{S}$ feed back into the mode dynamics through the coupling $\tilde{H}_i$. Later we will see  how quantum fluctuations of $\mathcal{S}$ lead to switching between metastable states of the mode. We will formally assume that the energy of the dispersive coupling satisfies $|V|\ll \tau_{_S}^{-1}$, although the actual condition of relevance is $V^2\tau_{_S}\ll \tau_{_M}^{-1}$, as will be seen below.

\subsection{Dynamics of system $\mathcal{S}$}
\label{sec:system_dynamics}

To begin, consider the case where the mode is set to be in an eigenstate $\Ket{m}$ of the operator $\hat{M}$: $\hat{M}\Ket{m} = m\Ket{m}$. The joint system-mode density matrix is given by the tensor product $\rho = \rho_{_S}\otimes\Ket{m}\Bra{m}$.
If we neglect the slow mode dynamics generated by the Liouvillian $\hat{L}_{_M}$ in Eq.~(\ref{eq:rho_{_S}chematic}), 
the reduced density matrix $\rho_{_S}$ of $\mathcal{S}$ obeys
\begin{equation}
\label{eq:rho_s_dot}
\dot{\rho}_{_S} = \hat{\Lambda}_m\rho_{_S};\quad \hat{\Lambda}_m\hat{O} = \hat{L}_{_S}\hat{O} + i[\hat{O},\hat{H}_i(m)],
\end{equation}
where $\hat{O}$ and $\hat{H}_i(m)$ are operators acting only on $\mathcal{S}$. Because the dispersive coupling $\tilde{H}_i$ commutes with $\hat{M}$, here it acts on the spin variables through its projection onto the selected mode state $\Ket{m}$, $\hat{H}_i(m) = \Bra{m}\tilde{H}_i\Ket{m}$. We will use the solutions of Eq.~(\ref{eq:rho_s_dot}) as a basis to build up the solution to the full problem, Eq.~(\ref{eq:rho_{_S}chematic}).

We solve Eq.~(\ref{eq:rho_s_dot}) in terms of the eigenoperators $\{{{}}\chi^\alpha_m\}$ of the superoperator $\hat{\Lambda}_m$, which appears on its right hand side:
\begin{equation}
\label{eq:lambda_{_M}}
\hat{\Lambda}_m{{}}\chi^\alpha_m = -\lambda^\alpha_m {{}}\chi^\alpha_m.
\end{equation}
Note that, from Eq.~(\ref{eq:hermitivity}), $(\hat{\Lambda}_m{{}}\chi^\alpha_m)^\dagger = \hat\Lambda_m({{}}\chi^\alpha_m)^{\dagger}$. Therefore, if ${{}}\chi^\alpha_m$ is an eigenoperator of $\hat{\Lambda}_m$ with eigenvalue $-\lambda^\alpha_m$, then $({{}}\chi^\alpha_m)^\dagger$ is also an eigenoperator with the eigenvalue $(-\lambda^\alpha_m)^*$. The eigenoperators ${{}}\chi^\alpha_m$ with real eigenvalues can be chosen to be Hermitian.

If system $\mathcal S$ has $N_{_S}$ states ($N_{_S}=2$ for a TLS), operators ${{}}\chi^{\alpha}_m$ are $N_{_S}\times N_{_S}$ matrices. Because the superoperator $\hat{\Lambda}_m{{}}$ in Eq.~(\ref{eq:lambda_{_M}}) does not commute with its adjoint,
there is no guaranty that the set of eigenvectors (operators) $\{{{}}\chi^{\alpha}_m\}$ forms a complete basis for system $\mathcal{S}$. Specifically, under fine-tuned conditions, Eq.~(\ref{eq:lambda_{_M}}) may have less than $N_{_S}^2$ linearly independent solutions and additional steps are needed to solve the dynamical problem (\ref{eq:rho_s_dot}). Here we will not treat such secular cases, assuming that the operator $\hat{\Lambda}_m{{}}$ is diagonalizable. This condition is generically satisfied for problems of physical interest including the specific examples considered below. Furthermore, we will not consider the other structurally unstable case where some of the eigenvalues $\lambda^\alpha_m$ coincide, as such degeneracy is lifted by an infinitesimally small change of the parameters of $\hat\Lambda_m$.  

Since Eq.~(\ref{eq:rho_s_dot}) describes relaxation of the system $S$, the eigenvalues $\lambda^{\alpha}_m$ have non-negative real parts. One of these eigenvalues (with $\alpha=0$, for concreteness) is equal to zero, which corresponds to the stationary state of system $\mathcal S$ for the mode in state $\Ket{m}$. The minimal value of Re~$\lambda^{\alpha>0}_m$ is the relaxation rate of system $\mathcal{S}$ for a given $\Ket{m}$. The relaxation time $\tau_{_S}$ is given by the maximal value of $[{\rm Re}~\lambda^{\alpha>0}_m]^{-1}$ calculated for the characteristic $m$.

We define the inner product of system-$\mathcal{S}$ operators $\hat O_1, \hat O_2$ as $\langle\hat{O}_1,\hat{O}_2\rangle = {\rm Tr}_{_S}[\hat{O}_1^\dagger\hat{O}_2]$, where Tr$_{_S}$ is taken over the states of system $\mathcal S$. The expression $\langle\hat{O}_1,\hat\Lambda_m\hat{O}_2\rangle = {\rm Tr}_{_S}[\hat{O}_1^\dagger\hat\Lambda_m\hat{O}_2]$ then defines how the superoperator $\hat\Lambda_m$ acts to the left (in this case, on the operator $\hat O_1^\dagger$); it also defines the adjoint superoperator $\hat\Lambda_m^\dagger$ through $(\hat O^\dagger\hat\Lambda_m)^\dagger = \hat\Lambda_m^\dagger O$.

``Left'' eigenoperators $\{\chi_{\alpha m}^\dagger\}$ of $\hat\Lambda_m$, which we denote with lowered indices, are defined through the equation 
\begin{equation}
\label{eq:left_eigenoperators}
{{}}\chi^\dagger_{\alpha m}\hat{\Lambda}_m = -\lambda_{\alpha m}{{}}\chi^\dagger_{\alpha m}.
\end{equation}
The left and right {\it eigenvalues} of $\hat\Lambda_m$  coincide: from Eqs.~(\ref{eq:lambda_{_M}}) and (\ref{eq:left_eigenoperators}), $\lambda_m^\alpha=\lambda_{\alpha m}$. However, the left and right {\it eigenoperators} are not Hermitian conjugate. The non-degeneracy of the spectrum implies the orthonormality relation
\begin{equation}
\label{eq:orthonorm}
\langle{{}}\chi_{\alpha m}, {{}}\chi^\beta_{m}\rangle = {\rm Tr}_{_S}[{{}}\chi^\dagger_{\alpha m} {{}}\chi^\beta_{m}] = \delta_{\alpha\beta},
\end{equation}
where we have imposed an additional normalization condition ${\rm Tr}_{_S}[{{}}\chi_{\alpha m}^\dagger {{}}\chi^\alpha_{m}] = 1$.
The orthogonality relation (\ref{eq:orthonorm}) holds only for the eigenoperators corresponding to the {\it same} mode state $\Ket{m}$. This will be important below when we consider evolution with general mode states which are not diagonal in $m$.

Over its relaxation time, system $\mathcal{S}$ reaches a quasi-stationary state for the given mode-state $\Ket{m}$. 
The reduced density matrix of $\mathcal{S}$ in the stationary state is given by the right eigenoperator ${{}}\chi^0_m$, corresponding to the zero-eigenvalue $\lambda^0_m = 0$. Note that the trace-preserving property of evolution dictates that ${\rm Tr}_{_S}[\hat{\Lambda}_m\hat{O}] = 0$, for any $\hat{O}$. By inserting the identity operator $\hat{I}_{_S}$ to the left of $\hat{\Lambda}_m$, we see that $\hat{I}_{_S}$ is a left eigenvector of $\hat{\Lambda}_m$ with eigenvalue 0. Hence we set ${{}}\chi_{0m}^\dagger= \hat{I}_{_S}$, such that the orthonormality condition (\ref{eq:orthonorm}) gives Tr$_{_S}\chi_m^0 =1$.
This is a very useful property, which we will employ below.

\subsection{Dynamics  of mode $\mathcal{M}$}
\label{sec:mode_dynamics}

We now use the solutions of the previous section to build up the solution to the full problem of coupled dynamics. To begin, for each $\alpha$ we collect the set of eigenoperators $\{{{}}\chi^\alpha_m\}$, together with corresponding projectors onto the mode states, $\{\Ket{m}\Bra{m}\}$, to form a single operator
\begin{equation}
\label{eq:chi_alpha} 
\bm{}{}{\chi}^\alpha = \sum_m {{}}\chi^\alpha_m\otimes \Ket{m}\Bra{m},
\end{equation}
which acts on the variables of both $\mathcal{S}$ and $\mathcal{M}$. Similarly, we define ${{}}{\bm{}{\chi}}_\alpha^\dagger = \sum_m {{}}\chi_{\alpha m}^\dagger\otimes \Ket{m}\Bra{m}$. We write the full density operator as
\begin{equation}
\label{eq:expansion}
\rho = \rho_1 + \rho_1^\dagger,\qquad \dot\rho_1 = \hat L\rho_1,\qquad \rho_1= \sum_\alpha \bm{}{\chi}^\alpha{\bf} p_\alpha, 
\end{equation}
where the operator
\begin{equation}
\label{eq:p_alpha}{\bf} p_\alpha = \sum_{m,m'}p_\alpha^{mm'}\Ket{m}\Bra{m'}
\end{equation}
acts only on the mode $\mathcal{M}$. Note that $\bm{}{\chi}^\alpha$ in Eq.~(\ref{eq:chi_alpha}) and ${\bf} p_\alpha$ in Eq.~(\ref{eq:p_alpha}) do not commute. Therefore the ordering of operators in the definition of $\rho_1$ in Eq.~(\ref{eq:expansion}) is important. In explicit form, we have
\begin{equation}
\label{eq:rho_1}
\rho_1 = \sum_{mm'}\sum_{\alpha}p^{mm'}_\alpha{{}}\chi^\alpha_m\otimes\Ket{m}\Bra{m'}.
\end{equation}

The set of complex parameters $\{p_\alpha^{mm'}\}$, for $\alpha = 0,\ldots, N_{_S}^2-1$ and $m,m' = 0, 1, 2, \ldots$, completely specifies the density matrix of the compound system. The asymmetry, that ${{}}\chi^\alpha_m$ appears in Eq.~(\ref{eq:rho_1}) while ${{}}\chi^\alpha_{m'}$ does not, is accounted for once $\rho_1$ is added to its conjugate in forming the full density matrix $\rho$.

For $\rho_1$ of the form (\ref{eq:rho_1}), the density matrix depends on time through the coefficients $\{p^{mm'}_\alpha(t)\}$.
This parametrization proves to be convenient for the analysis of the slow dynamics of the mode. Thus, below we recast the master equation (\ref{eq:rho_{_S}chematic}) as a coupled set of equations for the operators $\{{\bf} p_\alpha(t)\}$.

The dynamical equation for ${\bf} p_\alpha$ is obtained by substituting $\rho_1$ in the form (\ref{eq:rho_1}) into the equation $\dot\rho_1=\hat L \rho_1$. We then project out the ${\bf} p_\alpha$ part by multiplying from the left by ${\bm{}{\chi}}_\alpha^\dagger$ and taking the trace over the variables of system $\mathcal{S}$. Using the relation
\begin{eqnarray}
\label{eq:auxiliary}
&&\Bra{m}\hat L_{_S}(\chi^\alpha p_\alpha) + i[\chi^\alpha p_\alpha,\tilde H_i]\Ket{m'}\nonumber\\
&&= (\hat\Lambda_m\chi^\alpha_m)p_\alpha^{mm'} + i\chi^\alpha_m\Bra{m}[p_\alpha,\tilde H_i]\Ket{m'}, 
\end{eqnarray}
which results from Eq.~(\ref{eq:rho_s_dot}), along with Eq.~(\ref{eq:lambda_{_M}}) for $\hat\Lambda_m\chi^\alpha_m$, we obtain:
\begin{eqnarray}
\label{eq:p_equation}
\!\!\!\!\!\!\!\!\!\!{\bf}\dot p_\alpha &=& -\bm{}{\lambda}^\alpha{\bf} p_\alpha + \hat{L}_{_M}{\bf} p_\alpha + i\sum_\beta \hat{\nu}_{\alpha\beta}{\bf} p_\beta - \sum_\beta \hat{k}_{\alpha\beta}{\bf} p_\beta,
\end{eqnarray}
with
\begin{eqnarray}
\label{eq:nu_and_k}
\nonumber \bm{}{\lambda}^\alpha &=& \sum_m \lambda^\alpha_m\Ket{m}\Bra{m},\\
\nonumber \hat\nu_{\alpha\beta}{\bf} p_\beta&=& {\rm Tr}_{_S}\left({\bm{}{\chi}}_\alpha^\dagger \bm{}{\chi}^\beta \left[{\bf} p_\beta, \tilde{H}_i\right]\right),\\
\hat k_{\alpha\beta}{\bf} p_\beta &=&\delta_{\alpha\beta}\hat {L}_{_M} {\bf} p_\alpha- {\rm Tr}_{_S}\left[{\bm{}{\chi}}_\alpha^\dagger \hat {L}_{_M}(\bm{}{\chi}^\beta {\bf} p_\beta)\right].
\end{eqnarray}
Here $\hat \nu_{\alpha\beta}$ and $\hat k_{\alpha\beta}$ are superoperators.  Both of them result from the dispersive $\mathcal M - \mathcal S$ coupling. For $\hat \nu_{\alpha\beta}$ this is obvious, as this term explicitly contains the coupling Hamiltonian $\tilde H_i$ and goes to zero where the coupling  energy $V\to 0$. The term $\propto\hat k_{\alpha\beta}$arises because the eigenoperators $\{\chi^\alpha_m\}$ characterizing the dynamics of system $\mathcal{S}$ depend on the mode state $m$, via the coupling. As a consequence, the superoperator $\hat L_{_M}$ that describes dissipation of mode $\mathcal M$ does not commute with $\chi^\alpha$, i.e.,  $\hat{L}_{_M}(\bm{}{\chi}^\beta {\bf} p_\beta) \neq \bm{}{\chi}^\beta\hat{L}_{_M} {\bf} p_\beta$. We note that, since  ${\bm{}{\chi}}_0^\dagger = \hat I_{_S}$, we have $\hat{k}_{0\beta}{\bf} p_\beta = 0$. Also, since $\hat k_{\alpha\beta}$ comes from the mode dissipation, its typical size is of order $\propto \tau_{_M}^{-1}$.

We are interested in the effective nonlinear dynamics of the mode, described by the evolution of its reduced density matrix $\rho_{_M} = {\rm Tr}_{_S}\rho$. Since ${\rm Tr}_{_S}{{}}\chi^\alpha_m = 0$ for all $\alpha \neq 0$, we have 
\begin{equation}
\label{eq:rho_{_M}}
\rho_{_M}(t)={\bf} p_0(t)+{\bf} p_0^\dagger(t).
\end{equation}
However, as seen in Eq.~(\ref{eq:p_equation}), the evolution of ${\bf} p_0$ is coupled to the behavior of all ${\bf} p_{\alpha>0}$.
Thus to find $\rho_{_M}$ we must examine the full set of coupled dynamical equations.

If, as we assume, the relaxation rate $\tau_{_S}^{-1}$ of system $\mathcal S$ is large compared to the mode relaxation rate $\tau_{_M}^{-1}$ and to the coupling parameter in $\tilde H_i$ (divided by $\hbar$), the time evolution of $p_0$, described by Eq.~(\ref{eq:p_equation}), is qualitatively different from the evolution of operators $p_{\alpha>0}$. The evolution of ${\bf} p_0$ is governed by the mode Liouvillian $\hat{L}_{_M}$ and $\tilde{H}_i$, and therefore relaxation of $p_0$ is characterized by time $\tau_{_M}$.  In contrast, the relaxation rate of ${\bf} p_\alpha$ for $\alpha \neq 0$ is determined by the values of ${\rm Re~}\lambda_m^\alpha \geq \tau_{_S}^{-1}$. Therefore, over time  $\tau_{_S}$, all operators ${\bf} p_{\alpha > 0}$ approach quasi-stationary solutions of Eq.~(\ref{eq:p_equation}) for $\alpha>0$, calculated for the instantaneous $ p_0$. Moreover,  the matrix elements $\{p^{mm'}_{\alpha > 0}\}$ become small compared to the matrix elements of $p_0$. This is because  $\hat k_{\alpha\beta}\propto \tau_{_M}^{-1}$ and $\hat\nu_{\alpha\beta}\propto V$, and as we assume ${\rm Re}[\lambda_m^{\alpha > 0}]$ is large compared to $\tau_{_M}^{-1}, |V|$; as we will see below, the actual constraint on $|V|$ is significantly weaker.

\section{Mean field approximation}
\label{sec:mean_field}

\subsection{Semiclassical approximation for the mode}
\label{subsec:semiclassical}

Operator $p_0$ is of primary interest, as it determines the density matrix of the mode (\ref{eq:rho_{_M}}). From the arguments of the previous section, for times $t\gg \tau_{_S}$, to the leading order in $\tau_{_S}/\tau_{_M}$, its time evolution is determined by equations
\begin{equation}
\label{eq:zero_order}
\hat\Lambda_m \chi_m^{0} = 0, \qquad \dot p_0 = \hat L_{_M} p_0 + i\hat \nu_{00}p_0.
\end{equation}
The physical picture behind Eq.~(\ref{eq:zero_order}) is that system ${\mathcal S}$ reaches quasi-equilibrium, with distribution $\chi_m^{0}$, for a given state $m$ of the mode, and then the mode (and the system) slowly evolve to the self-consistent stationary state given by equation $\hat L_{_M}p_0 + i\hat\nu_{00}p_0=0$. 

The superoperator $\hat \nu_{00}$, which describes the effect of the coupling to ${\mathcal S}$ on the mode dynamics, has a simple form. Indeed, $\chi_{0m}^\dagger=\hat I_{_S}$, whereas $\chi^0_m$ gives the stationary density matrix of system ${\mathcal S}$ for the mode being in state $\Ket{m}$. In particular, for two nearby mode states $\Ket{m}$ and $\Ket{m'}$, with $m,m' \gg 1$ and $|m - m'| \ll m$, Eqs.~(\ref{eq:chi_alpha}) and (\ref{eq:nu_and_k}) give to leading order in $(m-m')$:
\begin{eqnarray}
\nonumber \Bra{m}\hat\nu_{00}p_0 \Ket{m'} &\approx&  (m'-m)\nu(m)\, p_0^{mm'}\\
\label{eq:nu_m} \nu(m) &\equiv& \langle \partial_m\hat{H}_i(m)\rangle_{_S}, 
\end{eqnarray}
where $\langle{\hat O}(m)\rangle_{_S}\equiv {\rm Tr}_{_S}[\chi^0_m{\hat O}(m)]$ is the average over the stationary state of system ${\mathcal S}$ performed for the mode in a given state $m$ and $\partial_m\hat H_i(m)\approx \hat H_i(m+1)-\hat H_i(m)\approx \hat H_i(m)-\hat H_i(m-1)$. The quantity $\nu(m)$ characterizes the change of level spacing of the mode due to its coupling to system $\mathcal{S}$. This change affects the distribution over mode states $\Ket{m}$ in a driving field by tuning the mode closer or further away from resonance. In turn, this affects the distribution of the system $\chi^0_m$, which itself determines $\nu(m)$. It is this mechanism that leads to the multistability of the response in the mean-field approximation.  

We will assume that modulation of the mode is sufficiently strong that the mode is excited to states with $m\gg 1$. As we will check a posteriori, the characteristic width of the distribution over $m$ is then small compared to the characteristic $m$. It is convenient to change from $p_0^{mm'}$ and $\rho_{_M}^{mm'}=\Bra{m}p_0+p_0^\dagger\Ket{m'}$ to 
\begin{eqnarray}
\label{eq:semiclass_p_0}
p_0(\mu,\phi)&=&\sum_{m,m'}p_0^{mm'}\delta_{\mu,(m+m')/2}e^{i(m-m')\phi},\nonumber\\
\rho_{_M}(\mu,\phi)&=&\sum_{m,m'}\rho_{_M}^{mm'}\delta_{\mu,(m+m')/2}e^{i(m-m')\phi}.
\end{eqnarray}
In the considered case the ``center of mass'' parameter $\mu=(m+m')/2$ is large, $\mu\gg 1$, and the major contribution to $p_0(\mu,\phi)$ comes from terms with $|m-m'|\ll \mu$. In Eq.~(\ref{eq:zero_order}) for the matrix elements $p_0^{mm'}$ one can change to $p_0(\mu,\phi)$, with account taken of Eq.~(\ref{eq:semiclass_p_0}), and use the semiclassical approximation in which $\mu$ is quasicontinuous.  A similar analysis can be done for the operator $p_0^\dagger$. This allows calculating the mode density matrix $\rho_{_M}(\mu,\phi)$. 

From the normalization condition, in the semiclassical limit the equation for $\rho_{_M}$ should have a form of the continuity equation $\partial_t\rho_{_M}(\mu,\phi) = -\partial_\mu j_\mu -\partial_\phi j_\phi$. Here, ${\bf j}\equiv (j_\mu,j_\phi)$ is the probability current in variables $(\mu,\phi)$. In the approximation (\ref{eq:nu_m}) it is determined by the operator $\hat L_{_M}$ and $\nu(\mu)$, with $\mu \approx m$. Generally it has a drift part, which is proportional to $\rho_{_M}(\mu,\phi)$ but does not contain derivatives of $\rho_{_M}$, a diffusion part that contains first derivatives, and higher-order derivatives. The expansion in the order of the derivatives is an expansion in $1/\mu$, and moreover, in $\mu^{-1}\tau_{_S}/\tau_{_M}$, as will be also seen from the example below. We note that this is not the classical limit. The diffusion coefficient has a contribution from quantum fluctuations, which will be dominating in the example below. Clearly, the dynamics of system ${\mathcal S}$ is purely quantum.  

To the leading order in $1/\mu$, one should keep in ${\bf j}$ only terms $\propto \rho_{_M}$, i.e., ${\bf j}(\mu,\phi)\approx {\bf K}(\mu,\phi)\rho_{_M}(\mu,\phi)$. Vector ${\bf K}=(K_\mu, K_\phi)$ has the meaning of the force that drives the mode, in the rotating frame. With account taken of the explicit form of $\hat L_{_M}$ and $\hat \nu_{00}$, from Eqs.~(\ref{eq:zero_order}) - (\ref{eq:semiclass_p_0})
\begin{align}
\label{eq:current_0}
&K_\mu \approx -f_{\rm diss}(\mu) - F_{_M}\mu^{1/2}\sin\phi,\nonumber\\
&K_\phi\approx -(F_{_M}/2)\mu^{-1/2}\cos\phi  - \delta\omega_{_M} + \nu(\mu).
\end{align}
Function $f_{\rm diss}(\mu)\propto \tau_{_M}^{-1}$ describes the effect of dissipation of the mode. For dissipation given by the standard linear friction operator (\ref{eq:Liouville_osc}), $f_{\rm diss}(\mu)= 2\Gamma_{_M}\mu$. In a more general case the dependence on $\mu$ can be more complicated. It is important, however, that, since in our model the dissipation operator is independent of the modulation, the thermal reservoir on its own does not have a preferred vibration phase, and thus $f_{\rm diss}$ is independent of $\phi$. The reservoir coupling lead to diffusion over phase, however, in this section we do not consider the diffusion (it will be discussed later) and the corresponding terms are absent in Eq.~(\ref{eq:current_0}). 

\subsection{Stationary states}
\label{subsec:stationary_states}

The approximation ${\bf j} = {\bf K}\rho_{_M}$ corresponds to the mean-field approximation. The mean-field equations of motion for variables $\mu$ and $\phi$ are 
\begin{equation}
\label{eq:mean_field_explicit}
\dot \mu = K_\mu, \qquad \dot\phi = K_\phi.
\end{equation}
They can have stationary solutions ${\bf K}={\bf 0}$ which describe the stationary states of forced vibrations of the mode. In principle, equations of motion of the type (\ref{eq:mean_field_explicit}) could also have periodic solutions that correspond to periodic vibration in the rotating frame. However, such solutions require that $\n \cdot {\bf K} > 0$ at least somewhere in phase space. From Eq.~(\ref{eq:current_0}), $\n\cdot{\bf K}=- df_{\rm diss}/d\mu$ has the same form as in the absence of modulation, where the only stationary state is $\mu=0$, and therefore $\n \cdot {\bf K} < 0$. The positions of the stationary states $\mu_{\rm st}$ in the presence of modulation are given by 
\begin{align}
\label{eq:stationary_states}
&G(\mu_{\rm st})=
F_{_M}^2, \nonumber\\
G(\mu)&=\mu^{-1}f_{\rm diss}^2(\mu) + 4\mu[\nu(\mu)-\delta\omega_{_M}]^2.
\end{align}

From Eq.~(\ref{eq:mean_field_explicit}), the stationary state (\ref{eq:stationary_states}) is stable provided $dG/d\mu >0$. In the absence of modulation, the state $\mu=0$ (i.e., the zero-amplitude state) is stable on physical grounds  and thus the condition $dG/d\mu >0$ is always satisfied for small $\mu$; quite generally $f_{\rm diss}\propto \mu$ for $\mu\to 0$.  If function $G(\mu)$ is monotonic, Eq.~(\ref{eq:stationary_states}) has one solution and the mode has only one stable state of forced vibrations for all modulation amplitudes $F_{_M}$.
 
\subsection{Multistability of forced vibrations}
\label{subsec:multistability}

For nonmonotonic $G(\mu)$, the mode 
can have several stable 
vibrational
states for a given $F_{_M}$, i.e., it can display bi- or multi-stability. For large $\mu$, the function $G(\mu)$ is increasing with $\mu$, except for the nongeneric case where $|\delta\omega_{_M} - \nu(\mu)|$ decreases at least as fast as $\mu^{-1/2}$. Then, since $dG/d\mu>0$ both for small and large $\mu$, it can have only an even number of zeros. These zeros give the positions of the saddle-node bifurcation points $\mu_B$,
\begin{equation}
\label{eq:bif_point}
dG/d\mu = 0, \qquad \mu=\mu_B.
\end{equation}
As seen from Eqs.~(\ref{eq:stationary_states}) and (\ref{eq:bif_point}), if the modulation amplitude $F_{_M}$ is tuned to the bifurcational value $F_{_M}^{(B)}=G(\mu_B)^{1/2}$, for $\mu=\mu_B$  (and for the corresponding $\phi_B$ given by equation ${\bf K}={\bf 0}$)  stable and unstable stationary states $\dot\mu=\dot\phi=0$ merge. Thus, the number of coexisting stable states changes by one once $F_{_M}$ goes through $F_{_M}^{(B)}$. 
\begin{figure}[ht]
\includegraphics[width=8truecm]{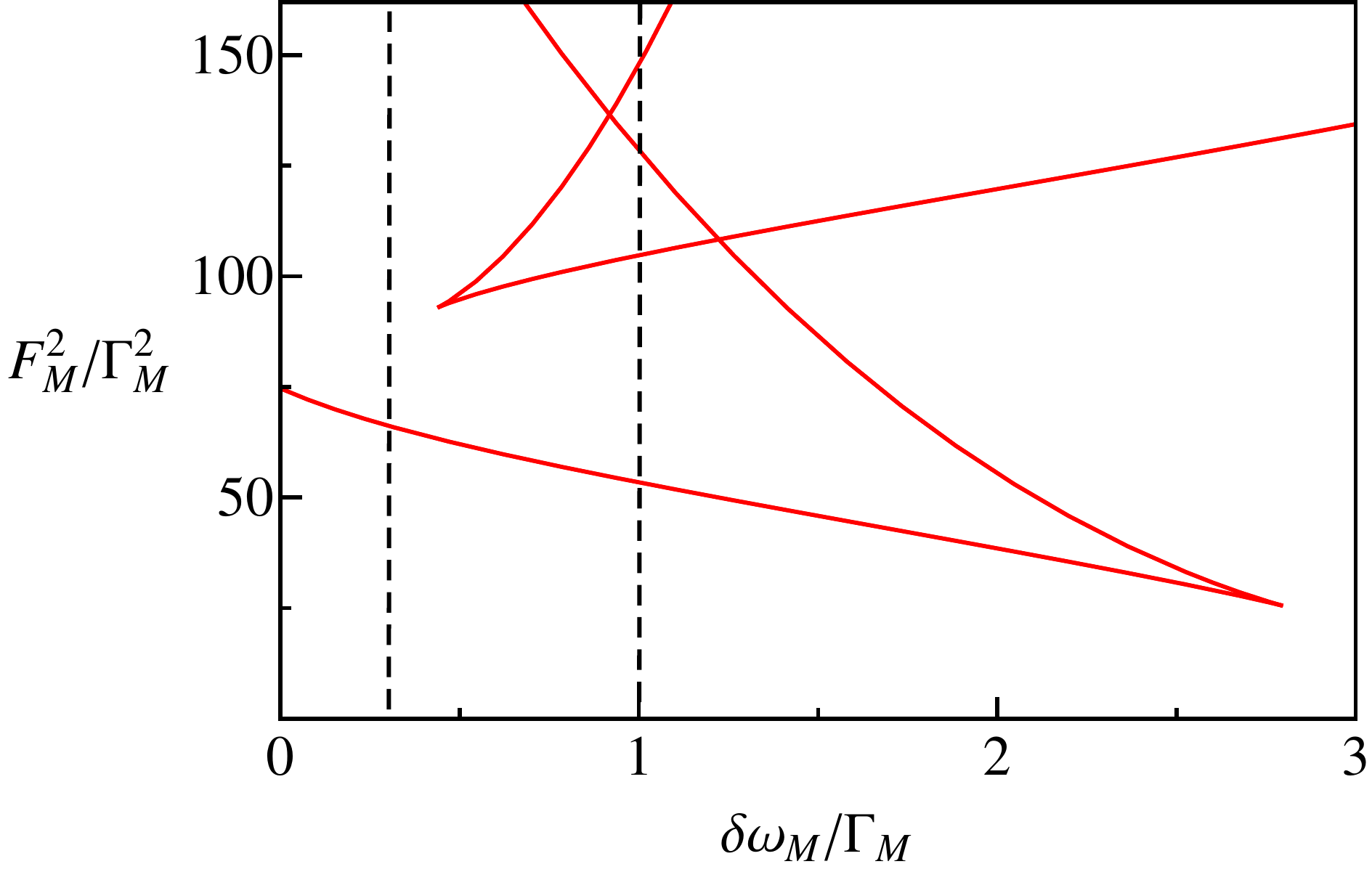}
\caption{The bifurcation diagram that shows the dependence of the bifurcation value of the modulating field $F_{_M}^{(B)}$ on the frequency detuning of this field $\delta\omega_{_M}=\omega_F-\omega_{_M}$ as given by Eqs.~(\ref{eq:stationary_states}) and (\ref{eq:bif_point}). The data refers to a mode dispersively coupled to a two-level system, with Hamiltonian (\ref{eq:H_RWA1}) and with dissipation described by linear friction, see Eq.~(\ref{eq:Liouville_osc}). The right dashed line shows the value of $\delta\omega_{_M}$ where the system displays tristability with varying $F_{_M}$; this $\delta\omega_{_M}$ corresponds to the main part of Fig.~\ref{fig:hysteresis}. The left dashed line shows $\delta\omega_{_M}$ used in the inset of Fig.~\ref{fig:hysteresis}. The other parameters are the same as in Fig.~\ref{fig:hysteresis}.}
\label{fig:bifurcation_diagram}
\end{figure}

The values of $\mu_{\rm st}$ in the stable states depend on the mode driving strength $F_{_M}$. Thus we consider the branches of stable states as functions of $F_{_M}$. These branches merge with the branches of unstable states at the bifurcation points (\ref{eq:bif_point}). The case where there are two stable-state branches and one branch of unstable states corresponds to vibration bistability and to the familiar $S$-shape dependence of the vibration amplitude  of the mode on the modulation amplitude, cf. \cite{LL_Mechanics2004}. For the model of a dispersively coupled mode and TLS discussed in Sec.~\ref{subsec:intro_multistability}, this dependence is shown in the inset of Fig.~\ref{fig:hysteresis}. In the region of bistability $G(\mu)$ has two extrema. If $G(\mu)$ has four extrema for a given set of modulation field parameters, the mode has three stable states, as seen in the main panel of Fig.~\ref{fig:hysteresis}.  

The bifurcational values $F_{_M}^{(B)}$ themselves depend on other parameters of the system, and in particular on the detuning of the modulation frequency $\delta\omega_{_M}$. The corresponding bifurcation curves are shown in Fig.~\ref{fig:bifurcation_diagram}. Each time any of these curves is crossed by varying parameters ($F_{_M}$ or $\delta\omega_{_M}$), the number of the stable and unstable states changes by one.

The bifurcation curves form pairs, which emanate from cusp point where the curves meet \cite{ArnoldNewYork1988}. Such cusp points are analogous to the critical points on lines of first-order phase transitions. If we are close to a cusp point and go around it in the $(F_{_M},\delta\omega_{_M})$ plane, without crossing the bifurcation curves, the number of stationary states does not change. If on the other hand, we move between the same initial and final values of $(F_{_M},\delta\omega_{_M})$ but cross the bifurcation curves that merge at the cusp point, we go through a region where there is an extra stable and an extra unstable state. On the bifurcation curves this unstable state must merge with two different stable states. At the cusp point all three states merge together.

The understanding of this topology makes the plot Fig.~\ref{fig:bifurcation_diagram} convenient. In particular, if we move up along the right dashed line, we start from one stable state for small $F_{_M}$, then there are added a stable and an unstable state once the lowest bifurcation curve $F_{_M}^{(B)}(\delta\omega_{_M})$ is crossed. When the next bifurcation curve is crossed, since it emanates from another cusp point, there is added another stable and unstable state. There are now three stable and two unstable states.  As we cross the still higher curve $F_{_M}^{(B)}(\delta\omega_{_M})$, the first unstable state merges with one of the stable states and disappears, so that the system now has two stable and one unstable state. When the highest curve $F_{_M}^{(B)}(\delta\omega_{_M})$ is crossed, there remains only one stable state. This behavior precisely corresponds to Fig.~\ref{fig:hysteresis}.

The ``beaks'' formed by the bifurcation curves in Fig.~\ref{fig:bifurcation_diagram}  open toward opposite sides. As follows from the above analysis, the tristability exists only in the range where the beaks overlap. In fact, the beaks do not go to infinity, they close up, but this occurs too far out to show on the figure.

\subsection{Slow dynamics near a bifurcation point}
\label{subsec:slow_dynamics}
We now consider the vicinity of a bifurcation point, i.e., we assume that $F_{_M}$ is close to $F_{_M}^{(B)}$, and expand the right-hand sides of the equations of motion (\ref{eq:current_0}) and (\ref{eq:mean_field_explicit}) about $\mu_B, \phi_B$. Here, the value of $\phi_B$ is given by the relation $K_\mu=K_\phi=0$, in which $\mu=\mu_B$ and $F_{_M}=F_{_M}^{(B)}$. If we limit the expansion of $K_\mu,K_\phi$ to linear terms in $\Delta\mu=\mu-\mu_B, \, \Delta\phi = \phi-\phi_B$, we find that one of the eigenvalues of equations (\ref{eq:mean_field_explicit}) for $\Delta\dot\mu, \Delta\dot\phi$ is equal to zero at the bifurcation point. Correspondingly, for the parameters close to the bifurcation point, a combination of the dynamical variables $\mu,\phi$ becomes ``slow,'' i.e.~there is a soft mode \cite{ArnoldNewYork1988}. The other eigenvalue of Eqs.~(\ref{eq:mean_field_explicit}) 
remains of order $\tau_{_M}^{-1}$ at the bifurcation point (note that we previously defined $\tau_{_M}$ as the relaxation time of mode ${\mathcal M}$ far from the bifurcation point). 

Over a time of order $\tau_{_M}$, the linear combination of $\Delta\mu, \Delta\phi$ corresponding to the large (negative) eigenvalue of Eq.~(\ref{eq:mean_field_explicit}) decays. After this decay, a relation between $\Delta\mu$ and $\Delta\phi$ is established, which to leading order can be obtained by linearizing equations (\ref{eq:mean_field_explicit}) in $\Delta\mu,\Delta\phi$ and setting $\Delta\dot\mu=\Delta\dot\phi=0$, with $F_{_M}=F_{_M}^{(B)}$. This gives $\Delta\phi=\xi_{\mu_B}\Delta\mu$, where $\xi_{\mu_B}=-\partial_\mu (f_{\rm diss}/\sqrt{\mu})/\{2\sqrt{\mu}[\nu(\mu)-\delta\omega_{_M}]\}$ with $\mu=\mu_B$. 

The slow dynamics near the bifurcation point are controlled by the quantity $Y=(2\mu)^{1/2}\sin\phi$, the soft mode, which 
happens to be the quadrature (out of phase) component of forced vibrations. At the bifurcation point, the deviation $\Delta Y=Y-Y_B$ of $Y$ from its bifurcational value $Y_B=(2\mu_B)^{1/2}\sin\phi_B$ is static, to linear order in $\Delta\mu, \Delta\phi$: $\Delta \dot{Y} = \partial_\mu Y \Delta\dot{\mu}\vert_{\mu = \mu_B} + \partial_\phi Y \Delta\dot{\phi}\vert_{{\mu = \mu_B}} = 0 + \mathcal{O}(\Delta\mu^2, \Delta\phi^2)$. Close to the bifurcation point, the soft mode dynamics are governed by the nonlinear equation
\begin{eqnarray}
\label{eq:slow_dynamics}
\Delta\dot Y=&&-\frac{\partial U}{\partial Y}, \qquad U(Y)=-\frac{1}{3}b_{}\Delta Y^3\nonumber\\
&&+(F_{_M}-F_{_M}^{(B)})\Delta Y/\sqrt{2},\nonumber\\
&&b_{}=(F_{_M}^{(B)}/8\sqrt{2})(\partial_\mu f_{\rm diss})^{-2}\partial^2_\mu G,
\end{eqnarray}
where the derivatives in the expression for $b_{}$ are calculated for $\mu=\mu_B$. As seen from Eq.~(\ref{eq:slow_dynamics}), if $b_{}(F_{_M}-F_{_M}^{(B)})>0$, the mode has a stable and an unstable stationary state. These states merge for $F_{_M}=F_{_M}^{(B)}$ and disappear for $F_{_M}$ on the opposite side of $F_{_M}^{(B)}$.

\section{Nonadiabatic fluctuations and switching between stable states}
\label{sec:fluctuations}

One of the best-known nonadiabatic effects in quantum systems is nonadiabatic transitions between stable states \cite{LL_QM81}. In the case we study here, nonadiabatic corrections to the mean-field theory also lead to transitions between the stable mode states. In the conventional picture, nonadiabatic transitions usually involve tunneling, for low temperature. In contrast, in our case nonadiabatic transitions are induced by fluctuations that come along with the relaxation \cite{Dykman2007}. Specifically, these are fluctuations due to the randomness of emission and absorption of excitations of the thermal reservoir by system ${\mathcal S}$. These fluctuations lead to fluctuations of the level spacing of the mode through the mode-system coupling. Classically, they correspond therefore to noise of the mode frequency. Even though the noise is of quantum origin, it causes activated-like interstate transitions over an effective barrier in phase space, see Eq.~(\ref{eq:switching_rate}).

The nonadiabaticity parameter is the ratio of the relaxation times $\tau_{_S}/\tau_{_M}$. Our analysis will be based on a perturbation theory. We will express functions $p_{\alpha>0}$ in the equation for the mode density matrix (\ref{eq:p_equation}) in terms of $p_0$ and then substitute them into equation (\ref{eq:zero_order}) for $p_0$. 

The major nonadiabatic corrections come from the term $\hat\nu_{\alpha\beta}p_\beta$ in Eq.~(\ref{eq:p_equation}); the terms $\hat k_{\alpha\beta}p_\beta$ are proportional to $\tau_{_M}^{-1}$ and thus lead to small corrections to the parameters of the operator $\hat L_{_M}$. To the leading order in  $\tau_{_S}/\tau_{_M}$, for time $t\gg\tau_{_S}$ one can set $\dot p_\alpha =0$ for $\alpha>0$, which gives a slowly varying in time solution $p_{\alpha>0}\approx i(\lambda^\alpha)^{-1}\hat\nu_{\alpha 0}p_0$. In turn, this gives an extra term in Eq.~(\ref{eq:zero_order}) for $p_0$, which now reads $\dot p_0 = \hat L_{_M} p_0 + i\hat\nu_{00}p_0 + \hat{\cal D}p_0$ with
\begin{eqnarray}
\label{eq:p_0_w_diffusion} 
\hat{\cal D}p_0=-\sum_{\alpha>0}(\lambda^\alpha)^{-1}\hat\nu_{0\alpha}[\hat\nu_{\alpha 0}p_0]
\end{eqnarray}
(here the superoperator $\hat\nu_{0\alpha}$ acts on the operator inside the bracket). 

We will be interested in the matrix elements $\Bra{m}\hat{\cal D} p_0\Ket{m'}$ between the mode states $\Ket{m}$ and $\Ket{m'}$. The calculation is simplified by the fact that operators $\tilde H_i$, $\chi_\alpha^\dagger$, and $\chi^\beta$ which appear in $\hat\nu_{\alpha\beta}$ are all diagonal in $m$. We will consider the semiclassical region of large $m,m'\gg 1$ and $|m-m'|\ll \mu=(m+m')/2$. As used above, in this region one can assume that $\mu$, $m$ are quasicontinuous variables and expand the coupling Hamiltonian, $\hat H_i(m)-\hat H_i(m')\approx (m-m') \partial_\mu\hat H_i(\mu)$. Then to the leading order in $m-m'$ we have
\begin{eqnarray}
\label{eq:D_explicit}
\Bra{m}\hat{\cal D} p_0\Ket{m'}& \approx &-(m-m')^2p_0^{mm'}\sum_{\alpha>0}{\rm Tr}_{_S}\left(\chi^\alpha_\mu\partial_\mu\hat H_i\right)\nonumber\\
&&\times {\rm Tr}_{_S}\left(\chi_{\alpha\mu}^\dagger \chi^0_\mu\partial_\mu \hat H_i\right)/\lambda^\alpha.
\end{eqnarray}

In addition to the leading order term displayed in Eq.~(\ref{eq:nu_m}), the function $\Bra{m}\hat\nu_{00} p_0\Ket{m'}$ in the equation for the matrix elements of $\dot{p}_0$ also has a term $\propto (m-m')^2$, i.e.,  
\begin{eqnarray*}
i\Bra{m}\hat\nu_{00} p_0\Ket{m'} &\approx&  i(m'-m)\nu(\mu)\,p_0^{mm'}\\ 
&&-\frac{i}{2}(m'-m)^2{\rm Tr}_{_S}[(\partial_\mu\chi^0_\mu)\,(\partial_\mu\hat H_i)]\,p_0^{mm'}.
\end{eqnarray*}  
It is helpful to further process the $(m'-m)^2$-term in this expression, by evaluating the quantity $\partial_\mu\chi^0_\mu$ which appears inside the trace.  This can be done by formally differentiating the equation $\hat\Lambda_m\chi^0_m=0$ over $m$, 
using Eq.~(\ref{eq:rho_s_dot}), and by expanding $\partial_m\chi^0_m$ in $\chi^\alpha_m$. The result is similar to the right-hand side of Eq.~(\ref{eq:D_explicit}), except for the extra factor $-(1/2)$ and the fact that in the second trace one should replace  the product of the operators $\chi^0_\mu$ and $\partial_\mu \hat H_i$ with their commutator. 

We note that an operator $\hat A(m)$ with respect to the variables of system ${\mathcal S}$ can be written as $\hat A(m)=\sum_\alpha\chi_{\alpha m}^\dagger\,{\rm Tr}_{_S}\left[ \chi^\alpha _m\hat A(m)\right]$. Further, it is convenient to consider system-${\mathcal S}$ operators in the Heisenberg representation in the rotating frame. From Eqs.~(\ref{eq:rho_s_dot}) and (\ref{eq:lambda_{_M}}), in this representation $\chi^\alpha_m(t)=\exp(-\lambda^\alpha_m t)\chi^\alpha_m(0)$ for $t\geq 0$; similarly, $\chi_{\alpha m}^\dagger(t)=\exp(-\lambda^\alpha_m t)\chi_{\alpha m}^\dagger(0)$. One can then define 
\[\hat A(m;t)= 
\sum\nolimits_\alpha\chi_{\alpha m}^\dagger \exp(-\lambda^\alpha_m t)\,{\rm Tr}_{_S}\left[ \chi^\alpha _m\hat A(m)\right].\]
Using this definition, one obtains 
\begin{equation*}
\label{eq:all_m-m}
\Bra{m}i\hat\nu_{00}p_0+\hat{\cal D}p_0\Ket{m'}\approx [i(m'-m)\nu(\mu)-(m'-m)^2D_\mu]p_0^{mm'}
\end{equation*}
with
\begin{align}
\label{eq:D_Time_correlator}
D_\mu=&{\rm Re}\int_0^\infty dt\left\langle\left[\partial_\mu\hat H_i(\mu;0) - \langle \partial_\mu\hat H_i(\mu)\rangle_{_S}\right]\right.\nonumber\\
 &\left.\times\left[\partial_\mu \hat H_i(\mu;t) - \langle \partial_\mu\hat H_i(\mu)\rangle_{_S}\right]\right\rangle_{_S}.
\end{align}
Here we used that, for real eigenvalues $\lambda^\alpha_m$, operators $\chi^\alpha_m, \chi^\dagger_{\alpha m}$ are Hermitian, whereas for the pairs of complex conjugate $\lambda^\alpha_m$ there are corresponding pairs of the Hermitian conjugate operators $\chi^\alpha_m, \chi^\dagger_{\alpha m}$. By its construction as the average of a correlator of the same operator over the states of system ${\mathcal S}$, the coefficient $D_\mu>0$. Clearly, $D_\mu$ is quadratic in the dispersive coupling constant $V$ contained in $\tilde H_i$, and  $D_\mu\propto\tau_{_S}$, i.e., $D_\mu\sim V^2\tau_{_S}$. Note that $D_\mu$, which we will see below plays the role of a diffusion constant, is small when the relaxation time of system $\mathcal{S}$ is very short. This dependence captures the motional narrowing that occurs when system $\mathcal{S}$ rapidly switches between its states.

\subsection{Diffusion equation for the  density matrix of the mode}
\label{subsec:symmetric}

With account taken of the terms $\propto (m-m')^2$ in equation for $p_0^{mm'}$, the equation for the density matrix $\rho_{_M}=p_0+p_0^\dagger$ in $(\mu,\phi)$-variables takes the form of the Fokker-Planck equation
\begin{equation}
\label{eq:rho_diffusion}
\dot \rho_{_M}=-\n\cdot({\bf K}\rho_{_M}) +D_\mu\partial^2_\phi\rho_{_M},\quad \rho_{_M}\equiv \rho_{_M}(\mu,\phi),
\end{equation}
where the vector $\n$ has components $\partial_\mu,\partial_\phi$ and the drift vector ${\bf K}$ is given by Eq.~(\ref{eq:current_0}). Function $\rho_{_M}$ satisfies the semiclassical normalization condition $(2\pi)^{-1}\int d\phi \, d\mu \rho_{_M}(\mu,\phi)=1$. 

We do not consider the diffusion term that comes from the direct coupling of the mode to the thermal reservoir, as described by the operator $\hat L_{_M}$. This term adds a contribution to the phase diffusion coefficient $D_\mu$ proportional to $\tau_{_M}^{-1}$; in addition, and importantly, this contribution is $\propto 1/\mu \ll 1$. Operator $\hat L_{_M}$ also introduces diffusion along the $\mu$-variable, with a diffusion coefficient that scales as $\tau_{_M}^{-1}/\mu$. Taking this diffusion into account will not change the analysis below, and in particular will just renormalize the coefficient $D_{\mu_B}$ in Eqs.~(\ref{eq:1D_diffusion}) and (\ref{eq:switching_rate}) below.

We assume that the diffusion is weak. This means that the distribution $\rho_{_M}$ in the stationary state has narrow peaks at the stable states of forced vibrations, which are given by the condition ${\bf K=0}$. From Eq.~(\ref{eq:rho_diffusion}), for $D_\mu\tau_{_M}\ll 1$, the peaks are Gaussian near the maximum. Their typical width is $(D_\mu\tau_{_M})^{1/2}$, and the peaks at different stable states are well separated from each other.

Further away from the stable states the stationary solution of Eq.~(\ref{eq:rho_diffusion}) can be sought in the eikonal form 
\begin{equation}
\label{eq:eikonal}
\rho_{_M}= \exp[-R(\mu,\phi)/D_\mu].
\end{equation}
To the leading order in $D_\mu$, function $R$ is independent of $D_\mu$ and can be found from a nonlinear equation of the form of the Hamilton-Jacobi equation \cite{DK_review84,Freidlin_book, Graham1984a}. 

One can see from the full nonadiabatic equation for the mode operators $p_\alpha$, Eq.~(\ref{eq:p_equation}), that the condition that the ratio $|p_{\alpha>0}/p_0|$ be small requires smallness of the parameter $|V|\tau_{_S} \overline{\Delta m}$, where $\overline{\Delta m}$ is the typical width of the distribution over $m$, or equivalently, $(D_\mu\tau_{_S})^{1/2} |\partial_\phi\ln p_0|\ll 1$. This estimate of $|p_{\alpha>0}/p_0|$ takes into account only the leading terms, which are described by the operator $\hat\nu_{\alpha\beta}p_\beta$, and applies in the time range $t\gg \tau_{_S}$ where all $p_{\alpha>0}$ have reached stationary values for a given $p_0$. From Eq.~(\ref{eq:eikonal}), $|\partial_\phi\ln p_0|\propto |\partial_\phi R|/D_\mu$. Near peaks of $\rho_{_M}$, where  $|\partial_\phi R|/D_\mu\lesssim(D_\mu\tau_{_M})^{-1/2} $, the condition $|p_{\alpha>0}/p_0|\ll 1$ reduces to $(\tau_{_S}/\tau_{_M})^{1/2} \ll 1$, which has been our major assumption all along.

On the far tail of the distribution we have $|\partial_\phi R|/D_\mu \sim (D_\mu\tau_{_M})^{-1} \gg 1$, and therefore the ratio 
$|p_{\alpha>0}/p_0|\propto (D_\mu\tau_{_M})^{-1/2}(\tau_{_S}/\tau_{_M})^{1/2}$ is not necessarily small. If this is the case, the adiabatic perturbation theory breaks down and the far tail of the distribution is not described by Eq.~(\ref{eq:rho_diffusion}). 
However, as we will see,  in the most interesting regime for studying the 
switching between metastable states, where the system is close to a bifurcation point, $|\partial_\phi R|\ll \tau_{_M}^{-1}$. 

\subsection{Switching rate near a bifurcation point}
\label{subsec:switching_bifurcation}

Equation (\ref{eq:rho_diffusion}) allows one to find, in a simple explicit form, the rate of switching from a metastable state near the saddle-node bifurcation point where this state disappears. Near this point the dynamics is controlled by the slow variable $Y(\mu,\phi)$, see Eq.~(\ref{eq:slow_dynamics}). The distribution $\rho_{_M}$ is a Gaussian peak with width $\sim (D_\mu\tau_{_M})^{1/2}$ in the direction transverse to the slow variable $Y$, whereas in the $Y$-direction it is much broader \cite{Dykman2007}. The distribution over the $Y$-variable  $\rho_{_M}(Y)=(2\pi)^{-1}\int d\mu\,d\phi\rho_{_M}(\mu,\phi)\delta\bigl(Y(\mu,\phi)-Y\bigr)$ can be found following the arguments of Ref.~\onlinecite{Dykman2007}. To the leading order in $D_\mu\tau_{_M}$ from Eqs.~(\ref{eq:slow_dynamics}) and (\ref{eq:rho_diffusion}) one obtains
\begin{equation}
\label{eq:1D_diffusion}
\dot\rho_{_M}(Y)=\partial_Y[\rho_{_M}(Y)\partial_Y U(Y)] 
+ D_{\mu B}\partial_Y^2\rho_{_M}(Y),
\end{equation}
where $D_{\mu, B}= D_{\mu}(\partial_\phi Y)^2_B$ is the coefficient of diffusion along the $Y$-axis, and $(\partial_\phi Y)_B=(2\mu_B)^{1/2}\cos\phi_B$ is the derivative of $Y(\mu,\phi) $ calculated at the bifurcation point. 

Equation (\ref{eq:1D_diffusion}) allows one to find the rate of escape, $W$, from a metastable state near a bifurcation point . This rate is described by the Kramers' theory \cite{Kramers1940},
\begin{equation}
\label{eq:switching_rate}
W= Ce^{-R_A/D_{\mu, B}}, \quad R_A=\frac{2^{5/4}}{3}\left[\frac{(F_{_M}-F_{_M}^{(B)})^3}{b_{}}\right]^{1/2}
\end{equation}
with $C=(2\pi)^{-1}\left[(F_{_M}-F_{_M}^{(B)})b_{}/\sqrt{2}\right]^{1/2}$.

As seen from Eq.~(\ref{eq:switching_rate}), the activation energy of switching near a saddle-node bifurcation point scales as the distance to the bifurcation point $F_{_M}-F_{_M}^{(B)}$ to the power 3/2. This is typical in the case where fluctuations are induced by Gaussian noise. In the present case this noise comes from quantum fluctuations of system ${\mathcal S}$ which modulate the frequency of the mode ${\mathcal M}$.

\subsection{Switching for coupling to a modulated two-level system}
\label{sec:TLS}

The analysis of Secs.~\ref{sec:adiabatic} - \ref{sec:fluctuations} can be applied, for example, to the problem of a mode coupled to a two-level system. A qualitative description of the mean-field dynamics of this model was given in Sec.~\ref{subsec:intro_multistability}. The consistent mean-field analysis outlined above leads to Eqs.~(\ref{eq:Bloch}) and (\ref{eq:naive_m}), with $m_{\rm st}$ replaced by the stationary value of the Wigner disitribution center-of-mass variable $\mu_{\rm st}$ of Eq.~(\ref{eq:stationary_states}). With this replacement, Eqs.~(\ref{eq:naive_m}) and (\ref{eq:stationary_states}) coincide. This justifies the results on the multistability of a mode coupled to a TLS presented in Sec.~\ref{sec:Introduction}.

\begin{figure}[ht]
\includegraphics[width=7.0truecm]{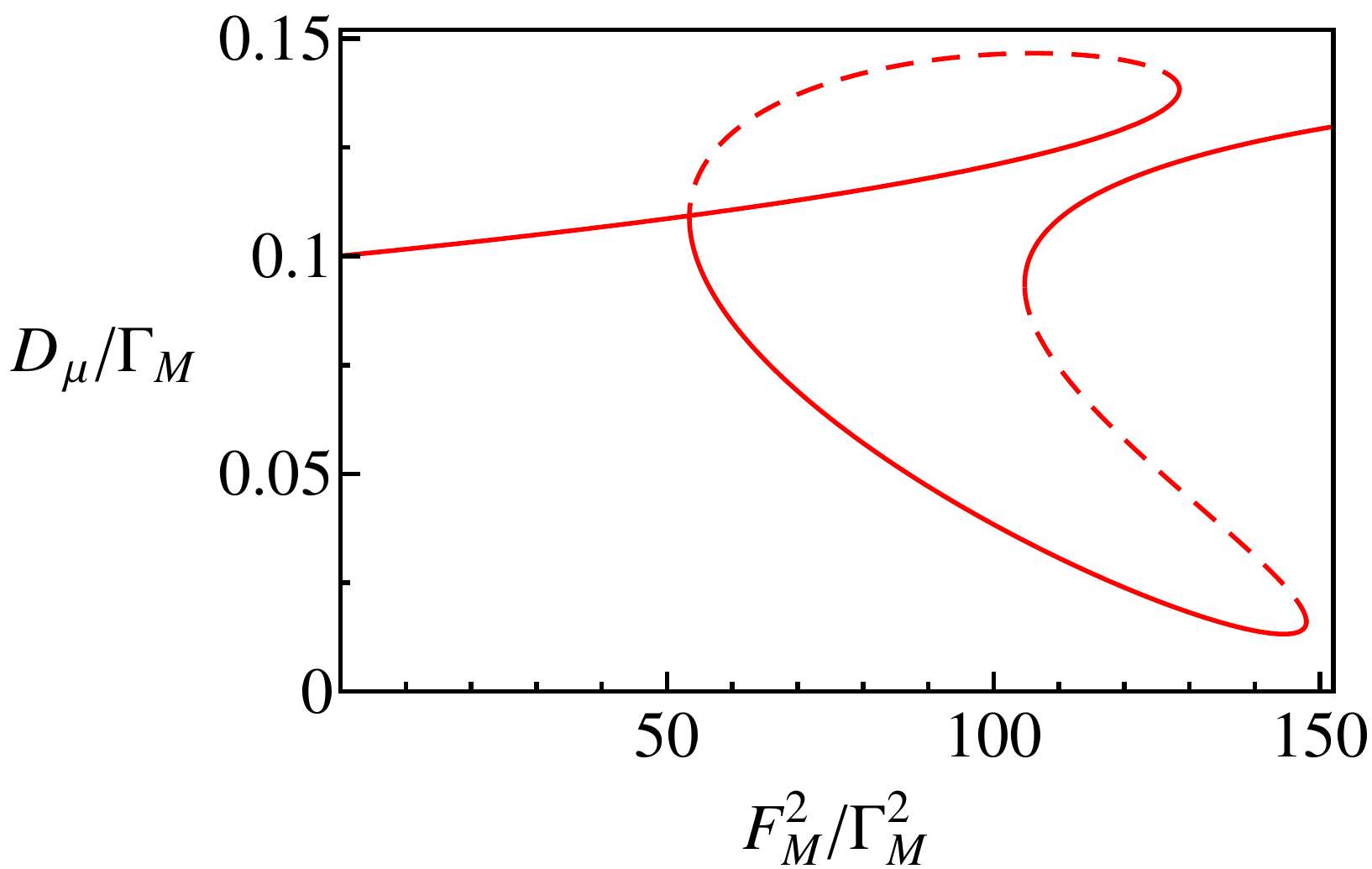}
\caption{The phase diffusion $D_\mu$ for a mode coupled to a TLS, scaled by the mode decay rate $\Gamma_{_M}$. The data refer to the mean-field characteristic in Fig.~\ref{fig:hysteresis}, which displays tristability.
The value of $D_\mu$ on the stable and unstable branches is shown by solid and dashed lines, respectively. 
Diffusion is caused by quantum fluctuations of the TLS, where we set the Planck number $\bar n(\omega_{_S})=0$. The point where the uppermost dashed line joins the lowermost solid line ($F_{_M}^2/\Gamma_{_M}^2\approx 54$) accidentally lies very close to the solid line that starts from $F_{_M}=0$ and corresponds to the lowest branch  $\mu_{\rm st}(F_{_M})$ in Fig.~\ref{fig:hysteresis}. } 
\label{fig:D_mu}
\end{figure}

The mean-field picture disregarded the effect of quantum fluctuations of the TLS. When relaxation of the mode is slow, the diffusion caused by these fluctuations is described by Eq.~(\ref{eq:rho_diffusion}). Using the Bloch equations for the TLS dynamics,
one can show that the effective diffusion coefficient of the mode's vibrational phase, which is defined by Eq.~(\ref{eq:D_Time_correlator}), has the form
\begin{eqnarray}
\label{eq:D_spin}
D_\mu &=& -V^2(\langle s_z\rangle_{_S}/4\Gamma_{_S})\left[1- 4\langle s_z\rangle_{_S}^2\right.\nonumber\\
&&\times\left.\left(1-\frac{1}{4}F_S^2
\frac{\gamma^2-\delta\omega_{_S}^2(\mu)}{[\gamma^2+\delta\omega_{_S}^2(\mu)]^2}\right)\right],
\end{eqnarray}
where $\langle s_z\rangle_{_S}$ is given by Eq.~(\ref{eq:Bloch}) with $m$ replaced by $\mu$. One can show from Eq.~(\ref{eq:Bloch}) that $D_\mu >0$. It is clear that $D_\mu\propto V^2\Gamma_{_S}^{-1}$ (see also discussion below Eq.~\ref{eq:D_Time_correlator}). The condition of the applicability of the approach is $D_\mu\tau_{_M}\sim V^2/\Gamma_{_M}\Gamma_{_S}\ll 1$. In Fig.~\ref{fig:D_mu} we show the scaled values of $D_\mu$ along the mean-field response curve of Fig.~\ref{fig:hysteresis} that displays tristability. As seen from this figure, $D_\mu/\Gamma_{_M}$ remains small for the considered example. 

We emphasize that for the TLS Planck number $\bar n_{_S}\rightarrow 0$, the noise described by the parameter $D_\mu$ is purely quantum. The noise is due to the randomness of spontaneous transitions between the states of the TLS, with corresponding emission of excitations of the thermal bath. On average, the transitions lead to relaxation of the TLS, but because they happen at random, they also cause fluctuations.

\section{Conclusions}
\label{sec:conclusions}

We have developed a theory of a vibrational mode (which may be of mechanical or electromagnetic origin) dispersively coupled to a quantum system, where both the mode and the system are driven far from thermal equilibrium. The results reveal new aspects of dispersive coupling. A profound consequence of the coupling is multistability of the nonlinear response, where the compound system can have multiple stable states in the mean-field approximation. This situation is very different from the familiar bistability due to intrinsic nonlinearity of a mode.

The multistability happens because, as a result of the interaction, the resonance frequency of the mode depends on the state of the system, while the state of the system depends on the degree of excitation of the mode. Effectively, the mode becomes nonlinear, with the transition frequency depending on the distribution over the states of the mode. In the simple but highly relevant case of the coupling to a two-level system, we found a regime where the compound system can have up to three stable states.

Our analysis refers to the case where the relaxation rate of the system is large compared to the mode relaxation rate. This case is of utmost interest for the broad range of currently studied compound systems. Here we discuss a few examples from the literature, which may naturally satisfy the conditions under which our description applies.

\begin{itemize}
\item {\it Example 1}: Double quantum dot charge qubit coupled to superconducting cavity mode. Such a system was studied in Ref.~\onlinecite{Petersson2012}, with mode frequency $\omega_{_M}/(2\pi) = 6.2$~GHz, mode lifetime $\tau_{_M} = 1 \times 10^{-7}$~s, qubit transition frequency $\omega_{_S}/(2\pi) = (2-7)$~GHz, qubit lifetime $\tau_{_S} = 1.5 \times 10^{-8}$~s, and Jaynes-Cummings coupling strength $g_c/(2\pi) = 30$~MHz. Here the cavity mode quality factor was only $Q = 2000$, already giving $\tau_{_M}/\tau_{_S} \gtrsim 1$. With expected device improvements the condition $\tau_{_M}/\tau_{_S} \gg 1$ will be reached. The dispersive coupling $V$ in our theory can be tuned via the qubit-cavity detuning $\omega_{_M} - \omega_{_S}$, giving e.g.~$V \sim g/100$, which easily satisfies the weak noise condition $V^2\tau_{_M}\tau_{_S} \ll 1$.
\item {\it Example 2}: Superconducting ``transmon'' qubit coupled to a superconducting stripline cavity mode. In the recent experiment described in Ref.~\onlinecite{Quintana2013}, the parameter values are: $\omega_{_M}/(2\pi) = 8.8$ GHz, $\tau_{_M} = 600$~ns, $\omega_{_S}/(2\pi) \approx 14$~GHz, $\tau_{_S} = 120$~ns, and Jaynes-Cummings coupling $g/(2\pi) \approx 180$~MHz. The condition $\tau_{_S}/\tau_{_M} < 1$ is weakly satisfied. The value of the dispersive coupling $V$ in our model is approximately equal to the ``qubit-qubit coupling'' $g_{12} \sim g^2/(\omega_{_S}-\omega_{_M}) \lesssim (2\pi)\times 10$~MHz. 
This gives $V^2\tau_{_M}\tau_{_S} \sim 10^2$, though the lifetimes and $V$ can be presumably decreased. Interestingly, this setup features two qubits coupled to the same cavity mode, allowing the possibility for studying the case where the system $\mathcal{S}$ has more than just two levels.
\item {\it Example 3}: Cooper pair box qubit coupled to a nanomechanical resonator. A device of this type was used in Ref.~\cite{LaHaye2009} to perform a nanomechanical measurement of the qubit state, with parameters: $\omega_{_M}/(2\pi) = 58$~MHz, $\tau_{_M}  \sim 100\ \mu$s, $\omega_{_S}/(2\pi) \sim 10$~GHz, $V/(2\pi) \sim 1$~kHz. The qubit relaxation time $T_1 \equiv \tau_{_S}$ was not measured, but from similar devices it can be expected to be of order $\tau_{_S} \sim 10$~ns.  Thus both the adiabaticity condition $\tau_{_S}/\tau_{_M} \approx 10^{-4} \ll 1$ and the weak noise limit $V^2\tau_{_M}\tau_{_S} \sim 10^{-5}$ are easily satisfied.
\end{itemize}

We have shown that quantum fluctuations in the system dispersively coupled to the mode cause switching between coexisting stable states. The switching rates are explicitly calculated in the most interesting region, i.e., near bifurcation point where metastable states disappear. We find that the effective switching activation energy displays power-law scaling with the distance to the bifurcation point, with exponent 3/2. This analysis holds in the regime of weak quantum noise, quantified by the parameter combination $V^2\tau_{_M}\tau_{_S} \ll 1$. 

Going beyond the limit described above, we also expect that the system displays interesting behavior where the condition that the quantum noise is weak is violated. In this case quantum noise leads to unusually large fluctuations between the areas centered near the mean-field stable states. Such behavior may be manifested in some of the systems described above. In other words, the system becomes an amplifier of  the nonequilibrium quantum noise. A detailed analysis of this effect is beyond the scope of the present paper, but is a worthwhile direction for future study.

\acknowledgements
MID acknowledges support from the ARO, grant W911NF-12-1-0235, and the Dynamics Enabled Frequency Sources
program of DARPA.

\bibliographystyle{apsrev4-1}

%

\end{document}